\documentclass[a4paper,twosides]{article}
\usepackage{xeCJK,fontspec,multicol,multirow,graphics,fancyhdr,epstopdf}
\usepackage{amsmath,amsfonts,amssymb,bm,upgreek,mathrsfs,ccmap,mathcomp}
\usepackage{subfigure,siunitx,graphicx,dsfont,braket,caption}
\usepackage[pagewise,switch,columnwise]{lineno}
\usepackage[compress,nospace]{cite}
\usepackage[dvipsnames]{xcolor}
\usepackage{CPL-2023}
\newcommand{\cplyear}{202x} \newcommand{\cplvol}{xx}
 \newcommand{\cplpagenumber}{xxxxxx}
 \newcommand{\cplpage}{\cplpagenumber-\thepage}

\begin{document}

\vspace* {-4mm} \begin{center}
\large\bf{\boldmath{Intervalley Band Crossing and Transition of Fractional Chern Insulators \\ in Floquet Twisted Bilayer MoTe$_2$}}
\footnotetext{\hspace*{-5.4mm}$^{*}$Corresponding authors. Email: zhaol@zju.edu.cn

\noindent\copyright\,{\cplyear}
\href{http://www.cps-net.org.cn}{Chinese Physical Society} and
\href{http://www.iop.org}{IOP Publishing Ltd}}
\\[5mm]

\normalsize \rm{}Yuhao Shi (石宇昊)$^{1}$, Zhao Liu (刘钊)$^{1,2*}$
\\[3mm]\small\sl $^{1}$Zhejiang Institute of Modern Physics, Zhejiang University, Hangzhou 310058, China
\\[3mm]\small\sl $^{2}$Zhejiang Key Laboratory of Micro-Nano Quantum Chips and Quantum Control, School of Physics, Zhejiang University, Hangzhou 310027, China
\end{center}

\vskip 1.5mm

\small{\narrower 
We study the twisted MoTe$_2$ homobilayer coupled to periodic driving of a circularly polarized light (CPL). Using Floquet theory in the high-frequency limit, we start from the Dirac model including both the valence and conduction bands of monolayer MoTe$_2$ to derive an effective time-independent Floquet Hamiltonian. The photon processes coupling the valence and conduction bands are captured in this Floquet analysis, and the resulting Floquet Hamiltonian contains explicit time-reversal symmetry breaking terms that are absent if conduction bands are integrated out from the beginning of the derivation. Based on the Floquet Hamiltonian, we find the increase of CPL driving intensity can cause the crossing of Floquet bands and redistribution of holes between the two valleys. When interactions are included, a transition between Floquet Laughlin-type FCIs with different behaviors of valley polarization is identified at total hole filling $5/3$.
\par}\vskip 5mm
\begin{multicols}{2}

{\it Introduction.}
Floquet systems have emerged as a versatile platform for engineering novel states of quantum matter. By subjecting materials to periodic driving, one can dynamically modify their band structures and effective interactions, enabling topological phases and dynamical transitions that are inaccessible in equilibrium\ucite{castro2022floquet,khemani2016phase,potter2016classification,yao2017discrete,potirniche2017floquet,lerose2019prethermal}. These effects are typically described by an effective static Floquet Hamiltonian, obtained through high-frequency expansions\ucite{eckardt2015highfrequency,bukov2015universal,mikami2016brillouinwigner} or more general non-perturbative approaches\ucite{rodriguez-vega2021lowfrequency,vajna2018replica,vogl2020effectivea,vogl2019analog,verdeny2013accurate,vogl2019flow}. Over the past decade, this framework has been widely applied to both weakly and strongly correlated systems, highlighting Floquet engineering as a promising route to control quantum phases\ucite{PhysRevB.79.081406,oka2019floquet,tsuji2008correlated,tsuji2009nonequilibrium}.

This framework has been naturally extended to moir\'e materials, which provide a fertile setting for exploring the interplay between strong correlations and topology. For example, circularly polarized light (CPL) shining vertically across the twisted multilayer graphene has been shown to cause nontrivial modifications of the band topology\ucite{oka2019floquet,vogl2020effectivea,katz2020optically,topp2019topological,li2020floquetengineered,ikeda2020highorder,vogl2020floquet,rodriguez-vega2020floquet,PhysRevB.104.195429,PhysRevB.103.195146}, which can potentially support fractional Chern insulators (FCIs)\ucite{sun2011nearly,tang2011hightemperature,neupert2011fractional,sheng2011fractional,regnault2011fractional,parameswaran2013fractional,bergholtz2013topological,liu2024recent}, the lattice analogs of the celebrated fractional quantum Hall effect, in Floquet systems\ucite{grushin2014floquet,anisimovas2015role,hu2023floquet}. Advances have also been reported in twisted MoTe$_2$ homobilayers (tMoTe$_2$)\ucite{dong2024floquet,qin2024lightenhanced}, which host flat bands over a wide range of twist angles\ucite{wu2018hubbard,zhan2020tunability,devakul2021magic}. Unlike in twisted multilayer graphene, recent work focusing on the valence bands of tMoTe$_2$ found that the leading effect of the vertically applied high-frequency CPL was only a constant overall quasienergy shift of static bands\ucite{vogl2021floquet}. This shift is independent of valley and driving frequency $\Omega$. Nevertheless, topological transitions can be driven by a longitudinal light generated in a waveguide\ucite{vogl2021floquet}.

The valence and conduction band edges near the $K$ and $K'$ points of a monolayer MoTe$_2$ are well described by massive Dirac fermions\ucite{xiao2012coupled}. In previous works about tMoTe$_2$, this massive Dirac structure is mostly neglected by perturbatively dropping the conduction band which is separated from the valence band by a large gap, resulting in a free-electron description (parabolic dispersion) of the valence band edge of a monolayer MoTe$_2$. While this approximation works well for describing the band dispersion, it fails to capture the feature of time-reversal symmetry breaking in the original massive Dirac model\ucite{PhysRevResearch.4.L032024}, which may manifest itself under the driving of CPL. This observation motivates us to revisit the Floquet problem in tMoTe$_2$ using the full consideration of the massive Dirac structure in monolayers.
It is also interesting to ask if there exist Floquet FCIs induced by CPL in tMoTe$_2$ when interactions are included.

In this work, we investigate the off-resonant high-frequency Floquet engineering of tMoTe$_2$ by vertical CPL shining. Instead of adopting the free-electron model for each MoTe$_2$ monolayer from the beginning, we start from the massive Dirac model to derive the effective Floquet Hamiltonian, and integrate out the conduction band only at the end. This method allows us to capture the photon processes coupling the valence and conduction bands. In the obtained free-electron approximation of the Floquet Hamiltonian, we identify overall quasienergy shifts that depend not only on the driving frequency but also on the valley. This time-reversal symmetry breaking feature is absent  if one derives the Floquet Hamiltonian completely within the free-electron framework. We then incorporate the screened Coulomb interaction in the system, and confirm that it does not introduce additional interaction terms in the second-order ($1/\Omega^2$) corrections of the Floquet Hamiltonian that may affect the many-body physics of the driven system. By performing exact diagonalization of the many-body Floquet Hamiltonian at total hole filling $\nu_h=5/3$, we track the evolution of the ground state of the Floquet Hamiltonian when increasing CPL intensity drives the crossing of Floquet bands in opposite valleys, revealing the competition between Floquet Laughlin-type FCI phases with different behaviors of valley polarization.

{\it Dirac model of static tMoTe$_2$.}
We start our discussion from the single-particle Hamiltonian of the static tMoTe$_2$. We assume that the top and bottom layers are rotated by $\pm\theta/2$ from the AA stacked configuration. The moir\'e Hamiltonian in the framework of massive Dirac model takes the form of\ucite{xiao2012coupled,wu2019topological}
\begin{equation}\el{H_kin}
    \begin{aligned}
        \mathcal{H}_{\text{kin}}
        =\begin{pmatrix}
            h_{t,+} & T_+\\
            T_+^\dagger & h_{b,+}
        \end{pmatrix}\bigoplus 
        \begin{pmatrix}
            h_{t,-} & T_-\\
            T_-^\dagger & h_{b,-}
        \end{pmatrix},
    \end{aligned}
\end{equation}
where $h_{\ell,\xi}$ is the massive Dirac Hamiltonian of layer $\ell$ at the valley $\xi=\pm$ ($K$ and $K'$), and $T_\xi$ is the interlayer hopping matrix. Throughout this work, the layer index $\ell$ takes the values $t=+1$ and $b=-1$ for the top and bottom layers, respectively. 
The monolayer term $h_{\ell,\xi}$ includes both the conduction and valence band edges in valley $\xi$, as given by\ucite{xiao2012coupled} 
\begin{equation}
    \begin{aligned}h_{\ell,\xi}(\bm{k},\bm{r})
        &=\begin{pmatrix}
            \Delta_g+\Delta_{\ell,c}(\bm {r})&0\\
            0&\Delta_{\ell,v}(\bm{r})\end{pmatrix}\\
            &+e^{+i\ell\xi\frac\theta4\sigma_z}[\hbar v_F
            \delta \bm{k}_\xi^\ell\cdot
            (\xi\sigma_x,\sigma_y)]e^{-i\ell\xi\frac\theta4\sigma_z}
    \end{aligned},\el{dirac_massive}
\end{equation}
where $c$ and $v$ denote the conduction and valence bands, respectively, $\Delta_g\approx\SI{1.1}{eV}$ is the average band gap, and $v_F$ is the Fermi velocity. We assume that the electron's spin in the Dirac model is polarized to opposite directions in the two valleys (spin-valley locking), which is caused by the strong spin-orbit coupling (SOC) at the valence band edge and the absence of spin flipping effects at the conduction band edge\ucite{xiao2012coupled}. The relative momentum $\delta \bm{k}_\xi^\ell=(\bm{k}-\boldsymbol{\kappa}_{\ell,\xi})$, with $\boldsymbol{\kappa}_{\ell,\xi}$ being the valley $\xi$ of layer $\ell$. $\Delta_{\ell,p}$ measures the variation of the extrema energy in band $p=c,v$ as a function of position $\bm{r}$, and takes the form of
\begin{equation}
    \Delta_{\ell,p}(\bm{r})=2V_p\sum_{j=1}^{3}\cos(\bm{b}_{2j-1}\cdot\bm{r}-\ell\psi_p).
\end{equation}
Here $\bm{b}_{n}=\frac{4\pi}{\sqrt{3}a_0/\theta}R_z(\frac{n-1}{3}\pi)\hat{x}$ are the moire reciprocal lattice vectors, $R_z$ generates a counterclockwise rotation around the $z$-axis, and $a_0\approx\SI{3.52}{\angstrom}$ is the lattice constant of monolayer MoTe$_2$. The interlayer tunneling $T_\xi$ in valley $\xi$ is parametrized as
\begin{equation}\el{T}
    \begin{aligned}
        T_\xi(\bm{r})&=
        \begin{pmatrix}
            w_c & w_{cv}\\
            w_{vc} & w_v
        \end{pmatrix}\\
        &+\begin{pmatrix}
            w_c & w_{cv}e^{-i2\xi\pi/3}\\
            w_{vc}e^{i2\xi\pi/3} & w_v
        \end{pmatrix}
        e^{i \xi\bm{b}_{2}\cdot\bm{r}}\\
        &+\begin{pmatrix}
            w_c & w_{cv}e^{i2\xi\pi/3}\\
            w_{vc}e^{-i2\xi\pi/3} & w_v
        \end{pmatrix}
        e^{i \xi\bm{b}_{3}\cdot\bm{r}}.
    \end{aligned}
\end{equation}

{\it Floquet engineering.} In the following, we include the effect of the vertically applied CPL driving in the single-particle physics by applying the Peierls substitution $\bm{k} \to \bm{k} +e\bm{A}/\hbar$ with the vector potential $\bm{A}=A_0(\cos{\Omega t},-\sin{\Omega t})$ in Eq.~(\ref{edirac_massive}). The light field is represented by an in-plane electric field ${\bm E}=-\frac{\partial{\bm A}}{\partial t}$. Here $A_0$ measures the driving strength and $\Omega$ is the driving frequency. Then Eq.~(\ref{eH_kin}) is modified to a time-dependent single-particle Hamiltonian $\mathcal{H}_{\text{kin}}(t)$. Note that we keep the interlayer tunneling as in the static case, because it is dominated by hopping between atoms that are exactly on top of each other, thus mostly contributed by the $z$-component of the vector potential which is absent in our setup. Because the light driving does not flip spin, the electron's spin is still polarized in each valley after the CPL is applied.

According to the Floquet theory, the stroboscopic evolution of the system can be captured by a time-independent effective Floquet Hamiltonian $H_{\text{eff}}$ (upon a unitary transformation from micromotion)\ucite{rahav2003effective,goldman2014periodically,bukov2015universal,eckardt2015highfrequency}. At high frequencies, $H_{\text{eff}}$ can be represented by a series expansion of $1/\Omega$:
\begin{equation}\el{H_eff}
    H_{\text{eff}} \approx H_{\text{eff}}^{(0)} + H_{\text{eff}}^{(1)} + H_{\text{eff}}^{(2)},
\end{equation}
where we keep to the second order. We use $H_m=\frac{1}{T}\int_0^T \mathcal{H}_{\text{kin}}(t) e^{-i m \Omega t} dt$ to denote the Fourier component of $\mathcal{H}_{\text{kin}}(t)$. In our setup, $H_m$ is nonzero only if $m=0,\pm 1$. Using the Magnus expansion\ucite{rahav2003effective,goldman2014periodically,bukov2015universal,eckardt2015highfrequency}, we obtain
\vskip -4mm
\begin{subequations}
    \begin{align}
        H_{\text{eff}}^{(0)}&=H_0=\mathcal{H}_{\text{kin}},\el{Heff0}\\
        H_{\text{eff}}^{(1)}&=\frac{1}{\hbar\Omega}[H_1,H_{-1}]
   =\frac{(e v_F A_0)^2}{\hbar \Omega}
        \sigma_z\otimes \mathds{1}_2^\ell \otimes \xi_z, \el{Heff1}\\
        H_{\text{eff}}^{(2)}&=\frac{1}{2(\hbar\Omega)^2}[H_1,[H_0,H_{-1}]]+h.c.\notag \\
        &=\frac{(e v_F A_0)^2}{(\hbar\Omega)^2}\bigg[
            -\mathcal{H}_{\text{kin}} 
            +\frac{\Delta_{\ell,v}}{2}(\mathds{1}_2^p+\sigma_z)\otimes \mathds{1}_2^\ell \otimes \mathds{1}_2^\xi \notag \\
            &+\frac{\Delta_g+\Delta_{\ell,c}}{2}(\mathds{1}_2^p-\sigma_z)\otimes \mathds{1}_2^\ell \otimes \mathds{1}_2^\xi \notag \\
            &+T^\prime \otimes \frac{\gamma_x+i\gamma_y}{2}\otimes \mathds{1}_2^\xi 
            +(T^\prime)^\dagger \otimes \frac{\gamma_x-i\gamma_y}{2}\otimes \mathds{1}_2^\xi 
            \bigg]
            \el{Heff2},
    \end{align}
\end{subequations}
where $\mathds{1}_2^p$, $\mathds{1}_2^\ell$ and $\mathds{1}_2^\xi$ are $2\times 2$ identity matrices in the band, layer and valley spaces, respectively, $\sigma_i$ and $\gamma_i$ ($i=x,y,z$) are Pauli matrices in the band and layer spaces, respectively, and 
$T_\xi^\prime(\bm{r})=\text{diag}(w_v e^{i\xi\theta},w_c e^{-i\xi\theta})
 (1+e^{i \xi\bm{b}_{2}\cdot\bm{r}}+
        e^{i \xi\bm{b}_{3}\cdot\bm{r}})
$. Combining these terms together, we have
\begin{equation}\el{H_eff_all}
    \begin{aligned}
        H_{\text{eff}}&=
        h_{\ell,\xi}^\prime(\bm{k},\bm{r})
        \otimes \mathds{1}_2^{\ell}
        \otimes \mathds{1}_2^\xi\\
        &+\left\{\Delta_{T}^\xi(\bm{r})\otimes \left(\frac{\gamma_x+i\gamma_y}{2}\right)
        +h.c.\right\} 
        \otimes \mathds{1}_2^\xi,
    \end{aligned}
\end{equation}
with 
\begin{subequations}
    \begin{align}
        h_{\ell,\xi}^\prime(\bm{k},\bm{r}&)=\left(1-\frac{\Delta}{\hbar\Omega}\right)\Bigg\{
        \begin{pmatrix}
            \Delta_g+\Delta_{\ell,c}(\bm{r})&0\\
            0&\Delta_{\ell,v}(\bm{r})
        \end{pmatrix}
        \notag\\
        &+e^{+i\ell\xi\frac\theta4\sigma_z}
        \left[\hbar v_F\delta \bm{k}_\xi^\ell\cdot(\xi\sigma_x,\sigma_y)\right]
        e^{-i\ell\xi\frac\theta4\sigma_z}\Bigg\}\notag\\
        +\Big(\xi-\frac{\Delta_g}{\hbar\Omega}&\Big)\Delta\cdot\sigma_z
        +\frac{\Delta}{\hbar\Omega}
        \begin{pmatrix}
            \Delta_g+\Delta_{\ell,v}(\bm{r}) & 0\\
            0 & \Delta_{\ell,c}(\bm{r})
        \end{pmatrix},\\
        \Delta_{T}^{\xi}(\bm{r})&=T_\xi(\bm{r})
        +\frac{\Delta}{\hbar\Omega}[T_\xi^\prime(\bm{r})-T_\xi(\bm{r})].
    \end{align}
\end{subequations}
Here we have used $\Delta=(e v_F A_0)^2/(\hbar\Omega)$ to denote the intensity of CPL driving. We will fix $\hbar\Omega=\SI{3}{eV}$ in the remainder of this paper. This driving frequency can be readily realized by ultraviolet light. In the Supplemental Material (SM), we will show that the light driving at this frequency is off-resonant and the high-frequency Magnus expansion truncated at the order of $1/\Omega^2$ indeed well captures the quasienergy bands. 

The Floquet Hamiltonian Eq.~(\ref{eH_eff}) keeps both the valence and conduction bands of the monolayer MoTe$_2$. For comparison with the undriven case in which the free-electron model involving the valence bands only was extensively used in previous studies, we perform the Schrieffer-Wolff transformation to Eq.~(\ref{eH_eff_all}) to integrate out the conduction bands. With some further simplifications, we finally obtain a Floquet Hamiltonian $\tilde{H}_{\rm eff}$ for the valence bands under the free-electron approximation, which takes a very similar form to that in the undriven case\ucite{wu2019topological}. The intralayer part of $\tilde{H}_{\rm eff}$ is 
\begin{equation}\el{h_eff_sw_sim_main}
    \begin{aligned}
        \tilde{h}_{\ell,\xi}(\bm{k},\bm{r})
        =-\frac{\hbar^2 {\delta \bm{k}_\xi^\ell}^2}{2m^{*}}
        +\Delta_{\ell,v}(\bm{r})+\left(-\xi+\frac{\Delta_g}{\hbar\Omega}\right)\Delta,
    \end{aligned}
\end{equation}
where the valley-dependent effective mass is $m^{*}=[\Delta_g+2(\xi-\frac{\Delta_g}{\hbar\Omega})\Delta]/(2v_F^2)$.
The interlayer tunneling term is
\begin{equation}\el{H_eff_SW_tunneling_sim_main}
    \tilde{T}_{\xi}(\bm{r})=w_v (1
    +e^{i \xi\bm{b}_{2}\cdot\bm{r}}
    +e^{i \xi\bm{b}_{3}\cdot\bm{r}}).
\end{equation}
The details of our derivation are presented in the SM. 

\vskip 4mm
\fl{fig:band_structure_DM}
\centerline{\includegraphics[width=0.45\textwidth]{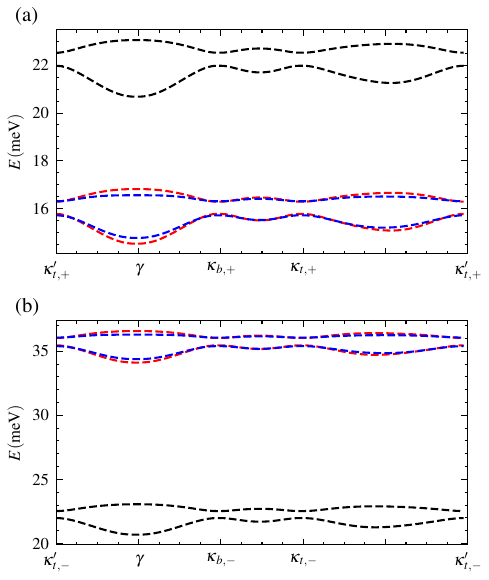}}
\vskip 2mm
\figcaption{8}{1}{The first two moiré valence bands in the two valleys along a high-symmetry path in the moiré Brillouin zone at a twist angle of \SI{1.2}{\degree}. 
(a) shows the bands in valley $+$, and (b) shows those in valley $-$. 
Black dashed lines represent the undriven tMoTe$_2$ described by the massive Dirac model. 
Red and blue dashed lines correspond to the quasienergies of the full Floquet Hamiltonian $H_{\text{eff}}$ and the valence-band Floquet Hamiltonian $\tilde{H}_{\rm eff}$, respectively. The intensity of CPL is chosen as $\Delta=\SI{10}{meV}$.}
\medskip

In Fig.~\fref{fig:band_structure_DM}{1}, we present the first two moir\'e valence bands in the two valleys, obtained for the static tMoTe$_2$, the full Floquet Hamiltonian $H_{\text{eff}}$, and the valence-band Floquet Hamiltonian $\tilde{H}_{\rm eff}$, using the parameters $\Delta_g=\SI{1.1}{eV}$, $v_F=\SI{0.4e6}{\m \per \s}$, $w_{cv}=\SI{15.3}{meV}$,
$(V_v,\psi_v,w_v)=(\SI{8}{meV},\SI{-89.6}{\degree},\SI{-8.5}{meV})$,
$(V_c,\psi_c,w_c)=(\SI{5.97}{meV},\SI{-87.9}{\degree},\SI{-2}{meV})$\ucite{wu2019topological}, and $\Delta=\SI{10}{meV}$. One can see that the quasienergy spectra of $H_{\text{eff}}$ and $\tilde{H}_{\rm eff}$ are indeed very similar to each other, which justifies our Schrieffer-Wolff transformation of integrating out the conduction bands. We will use the valence-band Floquet Hamiltonian $\tilde{H}_{\rm eff}$ in what follows.

From Fig.~\fref{fig:band_structure_DM}{1}, we can see that the dispersion of valence bands remains largely unchanged under periodic driving compared to the static case, except for an overall shift whose leading term $-\xi\Delta$ is in the order of $1/\Omega$ and is opposite in different valleys [Eq.~(\ref{eh_eff_sw_sim_main})]. This feature (as well as the valley-dependent effective mass) demonstrates the time-reversal symmetry breaking under the driving of CPL. Notably, this symmetry breaking signal is absent in the Floquet Hamiltonian derived within the free-electron framework\ucite{vogl2021floquet} (also shown in the SM), 
but could appear in the effective tight-binding description of the driven monolayer MoTe$_2$\ucite{dong2024floquet}. The necessity of including the conduction bands at the beginning of the Floquet analysis stems from the presence of photon processes coupling the valence and conduction bands, which can be seen in the nonzero matrix elements of $H_{\pm 1}$ between the valence and conduction band states (see the SM). If the static conduction bands are already integrated out before the Floquet analysis is performed, these processes cannot be captured.

The almost opposite quasienergy shift in the two valleys of driven tMoTe$_2$ immediately indicates that light can play the role of a pseudospin Zeeman field and cause the crossing between Floquet bands of different valleys. In Fig.~\fref{fig:schematic_plot_for_bilayer}{2}, we present the Floquet band structures at CPL intensity $\Delta=\SI{5}{meV}$, $\SI{10.3}{meV}$, and $\SI{15}{meV}$. With increasing $\Delta$, the top Floquet valence band of valley $+$ crosses with the second Floquet valence band of valley $-$. Such a crossing at the single-particle level will induce redistribution of electrons in the two valleys and could lead to interesting phase transitions of many-body physics in the driven system.   

\fl{fig:schematic_plot_for_bilayer}
\centerline{\includegraphics[width=0.45\textwidth]{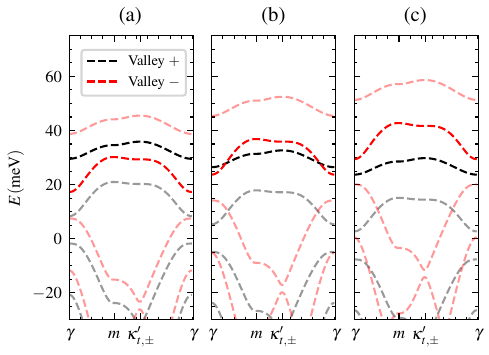}}
\figcaption{8}{2}{The Floquet band structures at CPL intensity (a) $\Delta=\SI{5}{meV}$, (b) $\SI{10.3}{meV}$, and (c) $\SI{15}{meV}$. The two crossing bands -- the top valence band of valley $+$ and the second valence band of valley $-$, are highlighted by dark colors.}    
\medskip

{\it Many-body Physics.} 
Now we will explore the many-body physics in driven tMoTe$_2$ that happens during the Floquet band crossing shown in Fig.~\fref{fig:schematic_plot_for_bilayer}{2}. The first step is to derive the many-body Floquet Hamiltonian with electron-electron interactions taken into account. For undriven tMoTe$_2$, the dual-gate screened Coulomb interaction
\begin{equation}\el{H_int}
    \mathcal{H}_{\text{int}}=\frac12\sum_{\bm{q}}{V(\bm{q}):\rho(\bm{q})\rho(\bm{-q}):}
\end{equation}
between electrons is often adopted, where $\rho(\bm{q})=\sum_j e^{i {\bm q}\cdot{\bm r}_j}$ is the density operator and $::$ means the normal ordering. The Fourier transform of the interaction potential is
\begin{equation}
    V(\bm{q})=\frac{e^2}{4\pi \epsilon_0 \epsilon_r}\frac{2\pi}{|\bm{q}|}\tanh{|\bm{q}|d},
\end{equation}
where $\epsilon_0$ is the vacuum permittivity, $\epsilon_r$ is the relative dielectric constant, and $d$ is the distance from a gate to the sample. The two valleys are decoupled at the single-electron level. Moreover, $ V(\bm{q})$ decays fast with increasing $|{\bm q}|$, so we neglect the intervalley scattering which requires large momentum transfer. Then we have conserved particle number in each valley and $\rho(\bm{q})=\sum_{\xi}{\rho_{\xi}(\bm{q})}$, where $\rho_{\xi}(\bm{q})$ is the density operator of electron in valley $\xi$. After the CPL driving is switched on, the interaction remains as in the undriven case since it has the density-density form. Therefore, regarding the Floquet Hamiltonian in the presence of the interaction, we need to add $\mathcal{H}_{\text{int}}$ to $H_{\rm eff}^{(0)}$ in Eq.~(\ref{eH_eff}), but $H_{\rm eff}^{(1)}$ is not altered. However, there is still a subtle thing: new terms could appear in $H_{\rm eff}^{(2)}$ due to the possible contribution from $[H_1,[\mathcal{H}_{\text{int}},H_{-1}]]$\ucite{anisimovas2015role}. Fortunately, after carefully calculating this commutator, we find it is exactly zero, as in driven TBG\ucite{hu2023floquet}. Then the total many-body Floquet Hamiltonian $\tilde{H}_{\rm eff}^{\rm mb}$ truncated at the order of $1/\Omega^2$ for the valence bands includes the single-particle part $\tilde{H}_{\rm eff}$ shown in Eqs.~(\ref{eh_eff_sw_sim_main}) and (\ref{eH_eff_SW_tunneling_sim_main}), and the interaction part $\mathcal{H}_{\text{int}}$ in Eq.~(\ref{eH_int}). 

Motivated by the recent breakthroughs of realizing FCIs in undriven tMoTe$_2$ at twist angles near $\SI{4}{\degree}$\ucite{park2023observation,zeng2023thermodynamic,cai2023signatures,xu2023observation}, we consider the possibility of Floquet FCIs resulting from the topology of Floquet bands and the electron-electron interactions, as well as the evolution of Floquet FCIs during the band crossing in Fig.~\fref{fig:schematic_plot_for_bilayer}{2}. Because we consider the off-resonant high-frequency light driving, we expect that the properties of the driven system can be well captured by the effective many-body Floquet Hamiltonian\ucite{PhysRevB.84.235108,grushin2014floquet,anisimovas2015role,hu2023floquet}. Therefore, we seek the signatures of Floquet FCIs from the eigenstates and spectrum of the effective many-body Floquet Hamiltonian using exact diagonalization. The specific occupation of Floquet bands required by Floquet FCIs could be realized by adiabatically increasing the CPL strength in an isolated Floquet system\ucite{PhysRevB.84.235108,dahlhaus2015magnetization}. We adopt the model parameters of tMoTe$_2$ provided by first-principles calculations in Ref.~\cite{wang2024fractional} for twist angles near $\SI{4}{\degree}$: $(V, w, \psi, m^*) = (\SI{20.8}{meV}, \SI{-23.8}{meV}, \SI{-107.7}{\degree},0.6m_e)$ and focus on $\theta=\SI{3.7}{\degree}$. For such a set of parameters, the Chern numbers of the top and second Floquet valence bands in valley $+$ are $+1$ and $-1$, respectively. The Chern numbers of Floquet valence bands in the other valley are opposite, as in the undriven case. We set $d=\SI{100}{\angstrom}$ and $\epsilon_r=10$, where signatures of static FCIs at hole filling $\nu_h=2/3$ were found in numerical simulations of undriven tMoTe$_2$\ucite{yu2024fractional}. 

Considering that the static FCIs in undriven tMoTe$_2$ were observed in the hole-doped valence bands, we will also work in the hole picture for the convenience of comparing undriven and driven systems. After the particle-hole transformation, we can express the entire many-body Floquet Hamiltonian in the basis $\ket{\bm{k},\xi,n}$ of hole bands as
\begin{eqnarray}\el{H_proj}
       \tilde{H}_{\rm eff}^{\rm mb}
        =-\sum_{\bm{k}}^{\text{MBZ}}\sum_{\xi}\sum_{n}
        E_{\xi,n}(\bm{k})\gamma_{\bm{k},\xi,n}^{\dagger}\gamma_{\bm{k},\xi,n}\nonumber\\
        +\sum_{\{\bm{k}_i\}}^{\text{MBZ}}\sum_{\xi,\xi^\prime }\sum_{\{n_i\}}
        V_{\{\bm{k}_i\}\{n_i\}}^{\xi\xi^\prime}
        \gamma_{\bm{k}_1,\xi,n_1}^{\dagger}\gamma_{\bm{k}_2,\xi^\prime,n_2}^{\dagger}
        \gamma_{\bm{k}_3,\xi^\prime,n_3}\gamma_{\bm{k}_4,\xi,n_4},
\end{eqnarray}
where all wave vectors are restricted in the moir\'e Brillouin zone (MBZ), $\gamma_{\bm{k},\xi,n}^{\dagger}(\gamma_{\bm{k},\xi,n})$ is the operator creating (annihilating) a hole with wave vector $\bm{k}$ in the Floquet valence band $n$ of valley $\xi$, and the quasienergy of holes is the negative of electron quasienergy $E_{\xi,n}(\bm{k})$ obtained by diagonalizing the single-particle part $\tilde{H}_{\rm eff}$.
The interaction matrix element 
\begin{eqnarray}
V_{\{\bm{k}_i\}\{n_i\}}^{\xi\xi^\prime}
=\frac{1}{2}\delta^\prime_{\bm{k}_1+\bm{k}_2,\bm{k}_3+\bm{k}_4}
\sum_{\bm{G}}V(\bm{k}_1-\bm{k}_4+\bm{G})\nonumber\\
\times M_{\xi,n_1,n_2}(\bm{k}_1,\bm{k}_4-\bm{G})
M_{\xi,n_2,n_3}(\bm{k}_2,\bm{k}_3+\bm{G}+\delta \bm{G}),
\end{eqnarray}
where $\delta^\prime$ is the periodic Kronecker delta function with the period of MBZ reciprocal lattice vector ${\bm G}$, $M_{\xi,n,n^\prime}(\bm{k},\bm{k}^\prime)=\langle u_{\xi,n}({\bm k})|u_{\xi,n'}({\bm k}')\rangle$ with $|u_{\xi,n}({\bm k})\rangle$ the periodic part of the eigenstate of $\tilde{H}_{\rm eff}$, and $\delta {\bm G}={\bm k}_1+{\bm k}_2-{\bm k}_3-{\bm k}_4$. 
\end{multicols}

\vskip 4mm
\fl{fig:energy_spectrum_noCPL}
\centerline{\includegraphics[width=\textwidth]{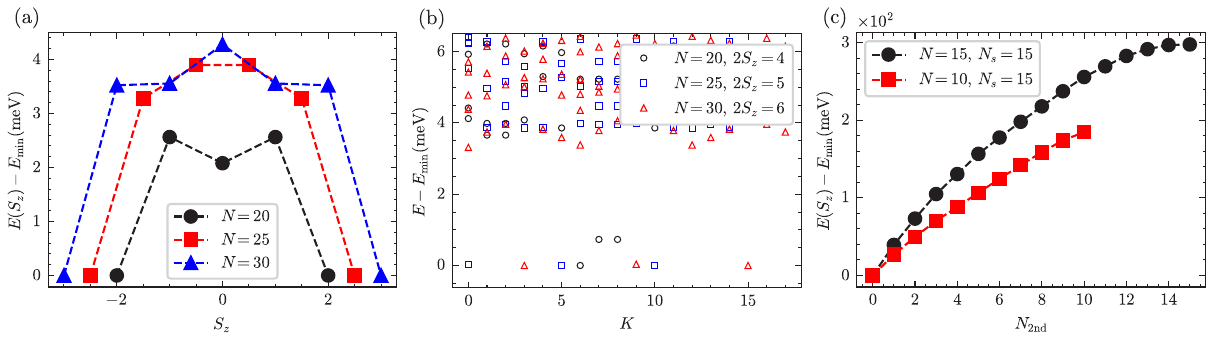}}
\figcaption{16}{3}{Results for undriven tMoTe$_2$ at $\nu_h=5/3$, with $N=20,25,30$ holes. (a) The lowest eigen-energy in each $S_z$ sector. (b) The many-body spectra in the $2S_z=4,5,6$ sectors. (c) Dependence of the ground energy at $\nu_h=1$ and $\nu_h=2/3$ on the redistribution of holes between the top two valence bands, with the valley polarization imposed.}

\begin{multicols}{2}
We will deal with finite periodic systems of $N$ holes on the torus, so that each energy level has a well-defined many-body momentum $K$. We choose the tilted geometry\ucite{läuchli2013hierarchy,repellin2014$mathbbz_2$} to make the samples as isotropic as possible. Because the number of holes $N_\xi$ in each valley $\xi$ is a good quantum number of Eq.~(\ref{eH_proj}), many-body eigenstates can be assigned with a $z$-direction pseudospin $S_z=(N_+-N_-)/2$ as well. We are interested in the many-body physics during the crossing between the Floquet bands in opposite valleys. Therefore, we consider the hole filling at $\nu_h=5/3$, where holes have to occupy both valleys. 

Before investigating the driven system, we would like to characterize the many-body phase at this filling in the undriven system ($\Delta=0$). For numerical efficiency, we first keep the top valence band in each valley and project the many-body Hamiltonian to these two bands. For all system sizes by us, we always find that the valley distribution of holes in the ground state is either $\nu_{h,+}=1,\nu_{h,-}=2/3$ or $\nu_{h,+}=2/3,\nu_{h,-}=1$, as shown in Fig.~\fref{fig:energy_spectrum_noCPL}{3}(a). These two energetically degenerate distributions correspond to $2S_z=\pm N/5$. The low-energy spectra in these $S_z$ sectors demonstrate approximate three-fold degeneracy [Fig.~\fref{fig:energy_spectrum_noCPL}{3}(b)], which is striking evidence of the Laughlin-type FCI. Notably, signatures of FCIs at $\nu_h=5/3$ were recently reported using transient optical spectroscopy\ucite{wang2025hidden}. Keeping two bands per valley exceeds our computational capability. However, for valley polarized undriven systems at either $\nu_h=1$ or $\nu_h=2/3$, we observe the ground energy increasing with $N_{\rm 2nd}$, the number of holes allowed to occupy the second valence band [Fig.~\fref{fig:energy_spectrum_noCPL}{3}(c)]. We hence argue that the mixing between the top two valence bands will not completely destroy the $\nu_h=5/3$ FCIs shown in Fig.~\fref{fig:energy_spectrum_noCPL}{3}(b). 

Next, we consider the driven tMoTe$_2$. After the CPL driving is turned on, the effective mass in $\tilde{H}_{\rm eff}$ in the two valleys becomes different. Moreover, the electron's Floquet bands gain a negative (positive) quasienergy shift in valley $\xi=+$ and $-$, respectively. Therefore, in the effective static system described by the many-body Floquet Hamiltonian $\tilde{H}_{\rm eff}^{\rm mb}$, the degeneracy at $\nu_h=5/3$ between the $2S_z=\pm N/5$ sectors in the undriven system is broken. For very small driving intensity $\Delta$, the Floquet FCI with $\nu_{h,+}=2/3,\nu_{h,-}=1$ configuration is selected as the ground state of $\tilde{H}_{\rm eff}^{\rm mb}$, because holes tend to occupy states with higher electron quasienergy. With the increase of $\Delta$, the quasienergies of the top Floquet valence band in valley $+$ become closer to that of the second Floquet valence band in valley $-$, as shown in Fig.~\fref{fig:schematic_plot_for_bilayer}{2}. In this case, holes may redistribute among the two valleys. 
\end{multicols}

\fl{fig:gs_sz_vs_delta}
\centerline{\includegraphics[width=\textwidth]{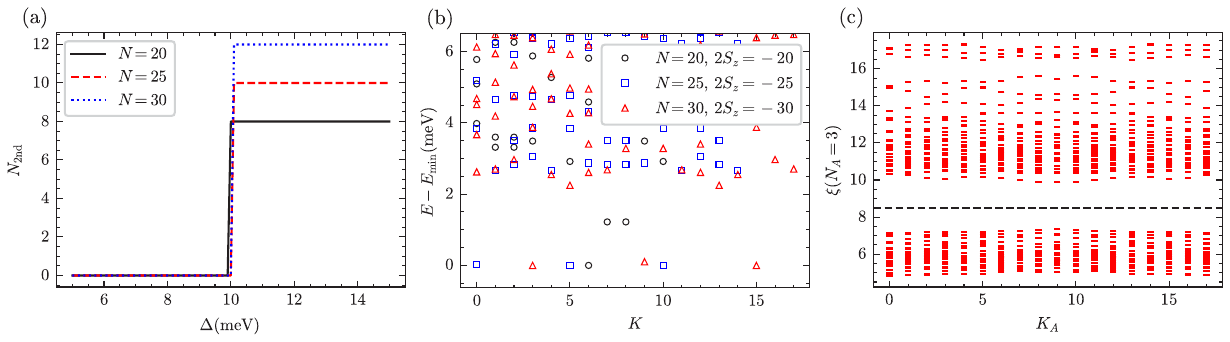}}
\figcaption{16}{4}{Results for CPL-driven tMoTe$_2$ at $\nu_h=5/3$, with $N=20,25,30$ holes. 
(a) The distribution of holes between the first Floquet valence band in valley $+$ 
and the second Floquet valence band in valley $-$ for different driving intensity $\Delta$. 
$N_{\text{2nd}}$ indicates the number of holes occupying the second Floquet valence band in valley $-$. (b) The low-energy spectrum at hole filling $5/3$ in valley $-$. 
We choose $\Delta=\SI{12}{meV}$. (c) The particle entanglement spectrum for $12$ holes in the second Floquet band of valley $-$ at $\nu_h=2/3$, after the particle-hole transformation and choosing $N_A=3$. There are 330 levels below the dashed line.}
\medskip

\begin{multicols}{2}

To study this redistribution of holes in the ground state of $\tilde{H}_{\rm eff}^{\rm mb}$, we assume that the holes only occupy the top valence band in each valley in the undriven limit. Furthermore, because the second Floquet valence band in valley $+$ and the top Floquet valence band in valley $-$ are well separated from the other two crossing bands, we assume that they are not relevant to the redistribution of holes induced by the CPL, namely, the hole fillings in these two bands remain at $0$ and $1$ during the band crossing, respectively. Then we reach a simplified model of band crossing, consisting of the top Floquet valence band in valley $+$ and the second Floquet valence band in valley $-$. These two bands are occupied at hole fillings $2/3$ and $0$ in the limit of $\Delta=0$, respectively. In Fig.~\fref{fig:gs_sz_vs_delta}{4}(a), we present the number of holes occupying the second valence band in valley $-$ as a function of $\Delta$ after the CPL driving is turned on. There is a clear transition at $\Delta_c\approx \SI{10}{meV}$, where all holes in the top Floquet valence band of valley $+$ move to the second Floquet valence band of valley $-$ (the number of holes occupying the top valence band in valley $+$ is $2N/5$ in the limit of $\Delta=0$). In this case, the distribution of holes becomes $\nu_{h,+}=0,\nu_{h,-}=5/3$ (the fully filled top Floquet valence band in valley $-$ is included). The occupations of holes in the mBZ before and after the transition are displayed in Fig.~\fref{fig:nk}{5}. 

After the transition, we project $ \tilde{H}_{\rm eff}^{\rm mb}$ to the top two Floquet valence bands in valley $-$. At hole filling $5/3$, we observe a three-fold ground-state degeneracy in the quasienergy spectrum, as shown in Fig.~\fref{fig:energy_spectrum&PES_for_2nd}{4}(b). We further compute the particle entanglement spectrum (PES)\ucite{regnault2011fractional} for the particle-hole conjugate of the three ground states projected to the second Floquet valence band in valley $-$. There is a clear gap in the PES, the number of levels below which matches the expectation for the $\nu=1/3$ Laughlin state [Fig.~\fref{fig:energy_spectrum&PES_for_2nd}{4}(c)]. These results point to the $\nu_h=1+2/3$ Laughlin-type Floquet FCI fully polarized in valley $-$, where the second Floquet valence band is partially occupied at $\nu_h=2/3$. By contrast, the Laughlin-type Floquet FCI at weak driving intensity ($\nu_{h,+}=2/3,\nu_{h,-}=1$) is not valley polarized, with hole occupations in both valleys. The valley polarized $\nu_h=5/3$ Floquet FCI at strong driving intensity does not have a counterpart in undriven systems because of the energy degeneracy of static bands in two valleys.
\vskip 4mm
\fl{fig:nk}
\centerline{\includegraphics[width=0.4\textwidth]{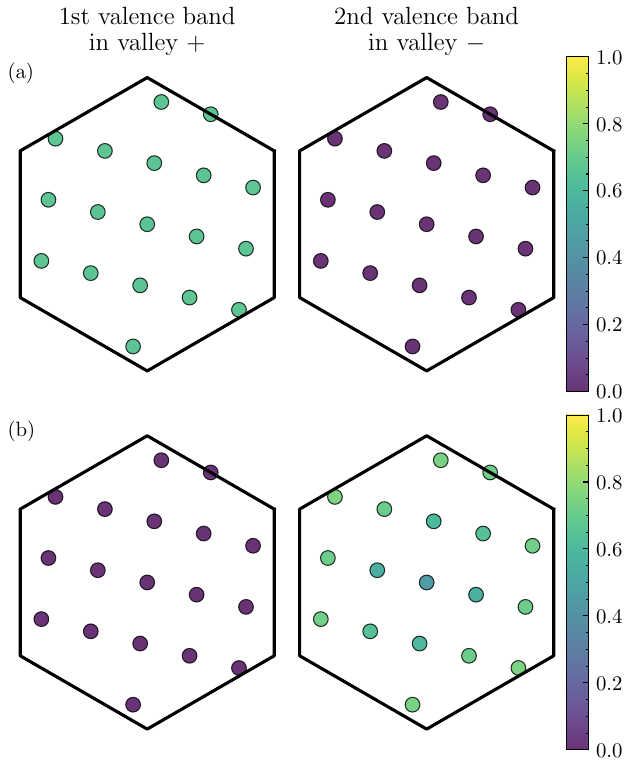}}
\figcaption{8}{5}{The hole occupation number in the mBZ for the first Floquet valence band in valley $+$ and the second Floquet valence band in valley $-$ when (a) $\Delta=\SI{5}{meV}$ and (b) $\Delta=\SI{12}{meV}$. The finite system includes 18 moir\'e unit cells.
\medskip}

{\it Conclusions and Discussions.}  
In this work, we have investigated the effect of circularly polarized light (CPL) on twisted bilayer MoTe$_2$. 
We derive the single-particle Floquet Hamiltonian in the high-frequency limit up to the order of $1/\Omega^2$, using the full Dirac model involving both valence and conduction bands of monolayer MoTe$_2$. 
This Floquet Hamiltonian can be further simplified by integrating out the conduction band via Schrieffer-Wolff transformation, resulting in a free-electron model where the time-reversal symmetry is explicitly broken at the order of $1/\Omega$. 
Such a symmetry breaking effect is hidden if one starts from a free-electron model at the beginning to derive the Floquet Hamiltonian. 
Therefore, our results highlight the necessity of including the conduction band contribution when studying the Floquet problems of tMoTe$_2$ and other semiconductor moir\'e materials even if the static band gap is large. 

We also explore the many-body physics of the driven tMoTe$_2$ at hole filling $\nu_h=5/3$ during the intervalley band crossing induced by the CPL. Using the effective many-body Floquet Hamiltonian, we find numerical evidence of Laughlin-type Floquet FCIs and their valley transition with increasing CPL strength. It is possible to realize the required occupation of Floquet bands by adiabatically increasing the CPL strength. The timescale needed to achieve such an adiabatic preparation could be very short (a few driving periods)\ucite{dahlhaus2015magnetization}, while the lifetime of the resulting prethermal steady state grows exponentially with increasing driving frequency\ucite{PhysRevB.95.014112}. For the high-frequency off-resonant driving focused in this work, we expect that the many-body physics extracted from the effective Floquet Hamiltonian well describes the properties of the non-equilibrium system. We note that the light-induced anomalous Hall effect, i.e., the Floquet integer Chern insulator has been observed experimentally\ucite{mciver2020lightinduced}. It will be exciting progress if the Floquet FCIs can be detected in experiments. 

In the future, it would be interesting to derive the effective Floquet Hamiltonian and study the many-body physics under the driving of a longitudinal light. It is also worth exploring the Floquet physics in other semiconductor moir\'e materials, like in twisted bilayer WSe$_2$.

{\it Acknowledgements.}  
This project was supported by the National Natural Science Foundation of China through Grant No.~12374149 and No.~12350403, and the National Key Research and Development Program of China through Grant No.~2020YFA0309200.

\bibliographystyle{iopart-num}
\bibliography{ref2}

\end{multicols}

\captionsetup[figure]{name=Fig.,labelfont=bf,labelsep=period}
\renewcommand{\thesection}{\small S\arabic{section}.}
\renewcommand{\thesubsection}{S\arabic{section}.\arabic{subsection}}

\numberwithin{equation}{section}
\renewcommand{\theequation}{S\arabic{section}.\arabic{equation}}
\counterwithin{figure}{section}
\renewcommand{\thefigure}{S\arabic{section}.\arabic{figure}}

\vspace* {-4mm} \begin{center}
\large\bf{\boldmath{Supplemental Material for: 
``Intervalley Band Crossing and Transition of Fractional Chern Insulators in Floquet Twisted Bilayer MoTe$_2$''}}
\end{center}
\vskip 5mm

This Supplemental Material includes three sections:
(1) The examination of the validity of the high-frequency expansion for the Floquet Hamiltonian of tMoTe$_2$ under circularly polarized light (CPL);
(2) Details for Schrieffer-Wolff transformation of the time-independent effective Floquet Hamiltonian;
(3) Derivation of the Floquet Hamiltonian completely within the free-electron framework.

\section{\small Examination of high-frequency expansion}
In the main text, we fix the CPL frequency at $\hbar\Omega = \SI{3}{eV}$. It is necessary to examine whether this value is sufficiently large so that the high-frequency expansion of the Floquet Hamiltonian truncated at the order of $1/\Omega^2$ captures the quasienergy spectrum of the driven tMoTe$_2$ well.
To this end, we compare the time-averaged density of states (DOS) obtained from the full time-dependent Hamiltonian $\mathcal{H}_{\text{kin}}(t)$ [Eq.~(1)] with the band structure of the effective time-independent Floquet Hamiltonian $H_{\text{eff}}$ [Eq.~(5)] introduced in the main text. 

According to the Floquet theory, the $\alpha$-th time-periodic Floquet mode $|u_\alpha(t)\rangle$ can be expanded as the Fourier series $|u_\alpha(t)\rangle=\sum_{m=-\infty}^{+\infty}e^{im\Omega t}|u^{\alpha}_{m}\rangle$. $|u^{\alpha}_{m}\rangle$ satisfies the time-independent eigenvalue equation
\begin{equation}
        \sum_{n=-\infty}^{\infty}\mathcal{H}_{mn}\ket{u^\alpha_{n}}
        =\varepsilon_\alpha \ket{u^\alpha_{m}},
    \end{equation}    
    where
    \begin{equation}
        \mathcal{H}_{mn}=m\hbar\Omega\delta_{mn}+
        \frac{\Omega}{2\pi}\int_{0}^{\frac{2\pi}{\Omega}} dt e^{-i(m-n)\Omega t}H(t)
\end{equation}
is the Fourier transform of the time-periodic Hamiltonian $H(t)$ and $\varepsilon_\alpha$ is the quasienergy of the system. 
In this work, the off-diagonal Fourier components, denoted as $H_{1}$ and $H_{-1}$ in the main text, are given by
\begin{subequations}
    \begin{align}
        H_{1}&=\xi e v_F A_0 e^{i\ell \theta/2}
        (\frac{\sigma_x+i\xi\sigma_y}{2})\otimes\mathds{1}_2^\ell\otimes\mathds{1}_2^\xi,\\
        H_{-1}&=H_{1}^{\dagger}
    \end{align}
\end{subequations}

The (momentum resolved) time-averaged DOS is defined as\ucite{katz2020optically}
\begin{equation}
    \el{DOS_time_avg}
    \bar{\rho}_{0}({\bm k},E)=\sum_{\alpha}\sum_{m}{A^{\alpha}_{{m}}({\bm k})
    \delta (\varepsilon _{\alpha}+m\hbar\Omega -E)},
\end{equation}
with $A^{\alpha}_{m}({\bm k})=\langle u^{\alpha}_{m}({\bm k})|u^{\alpha}_{m}({\bm k})\rangle$. Light driving may mix low-energy and high-energy bands if the driving frequency is smaller than the entire bandwidth of the system, which leads to smeared $ \bar{\rho}_{0}({\bm k},E)$. By contrast, sharp $ \bar{\rho}_{0}({\bm k},E)$ means that the driving frequency is sufficiently large to suppress the light-induced band mixing\ucite{katz2020optically}. 

\begin{figure}[!htp]
    \centering
    \includegraphics[width=\textwidth]{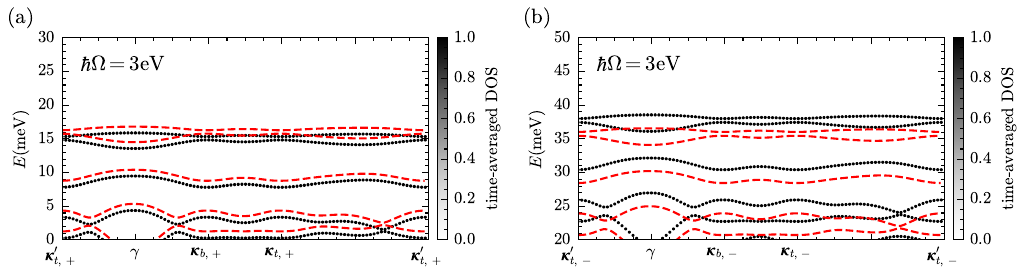}
    \caption{\label{fig:S1}Comparison of the time-averaged DOS of $\mathcal{H}_{\text{kin}}(t)$ (color map) with the band structure of $H_{\text{eff}}$ (red lines) at $\hbar\Omega = \SI{3}{eV}$. We truncate the Floquet modes $\ket{u^\alpha_{m}}$ to a finite number $N_F=2$, i.e., $m\in[-N_F,N_F]$. (a) is for the valley $\xi=+$, and (b) is for the valley $\xi=-$. The parameters are the same as those in Fig.~1 of the main text.}
\end{figure}

In Fig.~\ref{fig:S1}, we display the time-averaged DOS at $\hbar\Omega = \SI{3}{eV}$. Notably, $\bar{\rho}_{0}({\bm k},E)$ is very sharp, indicating the vanishing band mixing caused by light driving and the off-resonant feature of our driving. In this case, $\bar{\rho}_{0}({\bm k},E)$ also gives the quasienergy of the system, which is close to the band structure of the effective Floquet Hamiltonian $H_{\text{eff}}$ obtained by the high-frequency expansion (also shown in Fig.~\ref{fig:S1}). The energy bands in both cases have nearly the same dispersion shape, with discrepancies only up to $\sim \SI{2}{meV}$. Therefore, we confirm the validity of the high-frequency expansion at $\hbar\Omega = \SI{3}{eV}$.

\section{\small Details for Schrieffer-Wolff transformation}

In this section, we preform the Schrieffer-Wolff transformation to the time-independent effective Floquet Hamiltonian $H_{\rm eff}$ obtained with the Dirac model to 
integrate out the conduction bands. In the limit of large $\Delta_g$, we decompose the Floquet Hamiltonian as $H_{\text{eff}}=H^{(0)}+V$, where 
$H^{(0)}=\{\Delta_g /2\cdot\mathds{1}_2^p+[\Delta_g /2+(\xi-\frac{\Delta_g}{\hbar\Omega})\Delta]\cdot\sigma_z\}
\otimes\mathds{1}_2^\ell\otimes\mathds{1}_2^\xi$ with $H^{(0)}\ket{\ell,\xi,\zeta}=\tilde{h}^{(0)}_{\ell,\xi,\zeta}\ket{\ell,\xi,\zeta}$, and $V$ is treated as a perturbation. See the main text for the meaning of notations. 

Retaining terms up to second-order perturbation, we obtain the valence-band Floquet Hamiltonian $\tilde{H}_{\text{eff}}$ as 
\begin{equation}
    \el{H_eff_SW}
    \tilde{H}_{\text{eff}}=P\left(H^{(0)}+V_d+\frac{1}{2}[S,V_{od}]\right)P,
\end{equation}
where $S$ is determined by $[S,H]=-V_{od}$, and $P$ is the projection operator onto the valence band subspace. $V_d$ and $V_{od}$ are the diagonal and off-diagonal parts of $V$ with respect to the conduction and valence band subspaces.

By Eq.~(\eref{H_eff_SW}), the effective energy dispersion of the valence band in monolayer $\ell$ at valley $\xi$ is given by
\begin{equation}\el{h_eff_SW}
    \tilde{h}_{\ell,\xi,v}({\bm k},{\bm r})\approx
    \tilde{h}_{\ell,\xi,v}^{(0)}({\bm k},{\bm r})+
    \tilde{h}_{\ell,\xi,v}^{(1)}({\bm k},{\bm r})+
    \tilde{h}_{\ell,\xi,v}^{(2)}({\bm k},{\bm r}),
\end{equation}
with
\begin{subequations}
    \begin{align}
        \tilde{h}_{\ell,\xi,v}^{(0)}({\bm k},{\bm r})
        &=\left(-\xi+\frac{\Delta_g}{\hbar\Omega}\right)\Delta, \\
        \tilde{h}_{\ell,\xi,v}^{(1)}({\bm k},{\bm r})
        &=\bra{\ell,\xi,v}V_d\ket{\ell,\xi,v}\notag\\
        &=\left(1-\frac{\Delta}{\hbar\Omega}\right)\Delta_{\ell,v}({\bm r})
        +\frac{\Delta}{\hbar\Omega}\Delta_{\ell,c}({\bm r}),\\
        \tilde{h}_{\ell,\xi,v}^{(2)}({\bm k},{\bm r})
        &=\sum_{\ell^\prime}
        \frac{\mid\bra{\ell,\xi,v}V_{od}\ket{\ell^\prime,\xi,c}\mid^2}
        {\tilde{h}^{(0)}_{\ell,\xi,v}-\tilde{h}^{(0)}_{\ell^\prime,\xi,c}}\notag\\
        &=-\left(1-\frac{\Delta}{\hbar\Omega}\right)^2
        \left(\frac{\hbar^2 v_F^2 {\delta {\bm k}_\xi^\ell}^2}{\Delta_g^\prime}+ A_\ell\right),\\
        A_\ell=\frac{w_{cv}^2}{\Delta_g^\prime}&\left\{3+\sum_{j=1}^{3}\cos\left({\bm b}_{2j-1}\cdot{\bm r}-\ell\frac{2\pi}{3}\right)\right\},\\
        \Delta_g^\prime&=\Delta_g+2\left(\xi-\frac{\Delta_g}{\hbar\Omega}\right)\Delta.
    \end{align}
\end{subequations}
The interlayer tunneling part in $\tilde{H}_{\text{eff}}$ is given by
\begin{equation}\el{H_eff_SW_tunneling}
    \begin{aligned}
        \tilde{\Delta}_T^\xi({\bm r})&=\bra{t,\xi,v}V_d\ket{b,\xi,v}\\
        &+\sum_{\ell^\prime}
        \frac{\bra{t,\xi,v}V_{od}\ket{\ell^\prime,\xi,c}
        \bra{\ell^\prime,\xi,c}V_{od}\ket{b,\xi,v}}
        {\tilde{h}^{(0)}_{\ell,\xi,v}-\tilde{h}^{(0)}_{\ell^\prime,\xi,c}}\\
        &=\left\{w_v+\frac{\Delta}{\hbar\Omega}\left(w_c e^{-i\xi\theta}-w_v\right)\right\}
        \left(1+e^{i \xi{\bm b}_{2}\cdot{\bm r}}+
        e^{i \xi{\bm b}_{3}\cdot{\bm r}}\right)\\
        &-\left(1-\frac{\Delta}{\hbar\Omega}\right)^2\times
        \frac{\hbar v_F e^{-i\xi\frac{\theta}{2}}}{\Delta_g^\prime}\Big\{
            (\xi\delta k_x^t + i\delta k_y^t)(T_{\xi}({\bm r}))_{cv}\\
            &+(T_{\xi}({\bm r}))_{vc}(\xi\delta k_x^b - i\delta k_y^b)
        \Big\}.
    \end{aligned}
\end{equation}

Considering the limit of large $\Delta_g$ and $\hbar\Omega$, and the small $\Delta$ and $w_{cv}$, we further simplify Eq.~(\ref{eh_eff_SW}) and Eq.~(\ref{eH_eff_SW_tunneling}) to
\begin{equation}\el{h_eff_sw_sim}
    \begin{aligned}
        \tilde{h}_{\ell,\xi}({\bm k},{\bm r})&\approx
        -\frac{\hbar^2 v_F^2 {\delta{\bm k}_\xi^\ell}^2}{\Delta_g^\prime}
        +\Delta_{\ell,v}({\bm r})+\left(-\xi+\frac{\Delta_g}{\hbar\Omega}\right)\Delta\\
        &=-\frac{\hbar^2 {\delta {\bm k}_\xi^\ell}^2}{2m^{*}}
        +\Delta_{\ell,v}({\bm r})+\left(-\xi+\frac{\Delta_g}{\hbar\Omega}\right)\Delta
    \end{aligned}
\end{equation}
and 
\begin{equation}\el{H_eff_SW_tunneling_sim}
    \tilde{T}_{\xi}({\bm r})\approx w_v \left(1
    +e^{i \xi{\bm b}_{2}\cdot{\bm r}}
    +e^{i \xi{\bm b}_{3}\cdot{\bm r}}\right),
\end{equation}
respectively, with $m^{*}=[\Delta_g+2(\xi-\frac{\Delta_g}{\hbar\Omega})\Delta]/(2v_F^2)$.
Eqs.~(\ref{eh_eff_sw_sim}) and (\ref{eH_eff_SW_tunneling_sim}) give the $\tilde{H}_{\text{eff}}$ we use in the main text. The advantage of the simplification in Eqs.~(\ref{eh_eff_sw_sim}) and (\ref{eH_eff_SW_tunneling_sim}) is that the resulting $\tilde{H}_{\text{eff}}$ takes a very similar form to the undriven case, and only contains the parameters relevant to the valence bands which have been extensively investigated by first-principle calculations.

\begin{figure}[!htp]
    \centering
    \includegraphics[width=0.8\textwidth]{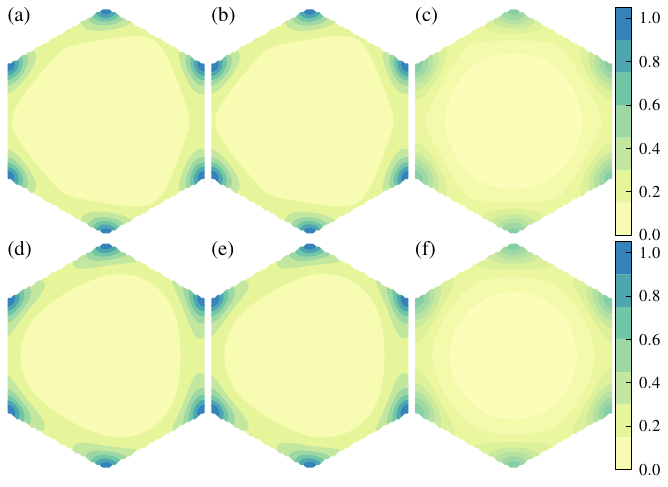}
    \caption{\label{fig:S2.1}Berry curvature 
    [(a)-(c)] and trace of quantum metric [(d)-(f)] for the top valence band in valley $\xi=+$. (a) and (d) are 
    for $H_{\text{eff}}$ obtained from the Dirac model; (b) and (e) are for the valence-band $\tilde{H}_{\text{eff}}$ without further simplification [Eqs.~(\ref{eh_eff_SW}) and (\ref{eH_eff_SW_tunneling})]; (c) and (f) are for the simplified $\tilde{H}_{\text{eff}}$ [Eqs.~(\ref{eh_eff_sw_sim}) and (\ref{eH_eff_SW_tunneling_sim})].
    }
\end{figure}

\begin{figure}[!htp]
    \centering
    \includegraphics[width=0.8\textwidth]{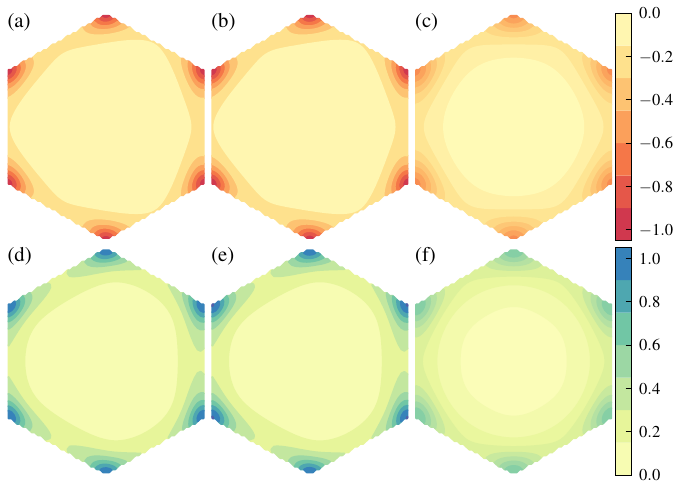}
    \caption{\label{fig:S2.2}Berry curvature 
    [(a)-(c)] and trace of quantum metric [(d)-(f)] for the second valence band in valley $\xi=+$. (a) and (d) are 
    for $H_{\text{eff}}$ obtained from the Dirac model; (b) and (e) are for the valence-band $\tilde{H}_{\text{eff}}$ without further simplification [Eqs.~(\ref{eh_eff_SW}) and (\ref{eH_eff_SW_tunneling})]; (c) and (f) are for the simplified $\tilde{H}_{\text{eff}}$ [Eqs.~(\ref{eh_eff_sw_sim}) and (\ref{eH_eff_SW_tunneling_sim})].}
\end{figure}

In the main text, we have shown that the band structures of $H_{\text{eff}}$ and $\tilde{H}_{\text{eff}}$ are very similar to each other. In Fig.~\ref{fig:S2.1} and \ref{fig:S2.2}, we further compare the Berry curvature and the trace of quantum metric for the 
top and the second valence bands at valley $\xi=+$, calculated from $H_{\text{eff}}$, as well as $\tilde{H}_{\text{eff}}$ with and without the simplification in Eqs.~(\ref{eh_eff_sw_sim}) and (\ref{eH_eff_SW_tunneling_sim}) [the latter corresponds to Eqs.~(\ref{eh_eff_SW}) and (\ref{eH_eff_SW_tunneling})]. 
The parameters are the same as those in Fig.~1 of the main text. The quantum geometry quantities share the same feature in all of the three cases, confirming the validity of both the Schrieffer-Wolff transformation and the further simplification in Eqs.~(\ref{eh_eff_sw_sim}) and (\ref{eH_eff_SW_tunneling_sim}).

\section{\small Floquet Hamiltonian within the free-electron framework}

The derivation of the Floquet Hamiltonian of tMoTe$_2$ under the CPL irradiation within the free-electron framework has been given in Ref.~\cite{vogl2021floquet}. In this section, we present the derivation for completeness.

Unlike in the main text where we start from the Dirac model to derive the Floquet Hamiltonian and integrate the conduction band only at the end, now we adopt the free-electron approximation from the beginning. The static moir\'e Hamiltonian in the free-electronic framework is given by
\begin{equation}
    \mathcal{H}_{\text{kin}}=
    \begin{pmatrix}
        h_{t,+} & T_{+}(\bm{r}) \\
        T_{+}^{\dagger}(\bm{r}) & h_{b,+}
    \end{pmatrix}\bigoplus
    \begin{pmatrix}
        h_{t,-} & T_{-}(\bm{r}) \\
        T_{-}^{\dagger}(\bm{r}) & h_{b,-}
    \end{pmatrix}
\end{equation}
with $h_{\ell,\xi}=-(\hbar \Delta \bm{k}_\xi^\ell)^2/(2m^*)+\Delta_{\ell,v}(\bm{r})$ and 
$T_{\xi}(\bm{r})=w_v(1+e^{i\xi\bm{b}_2\cdot\bm{r}}+e^{i\xi\bm{b}_3\cdot\bm{r}})$. Then we include the effect of the vertical applied CPL driving by applying the Peierls 
substitution $\bm{k}\rightarrow\bm{k}+e\bm{A}/\hbar$ with the vector potential 
$\bm{A}=A_0(\cos{\Omega t},-\sin{\Omega t})$.
The time-dependent monolayer Hamiltonian is
\begin{equation}
    \begin{aligned}
        h_{\ell,\xi}(\bm{k},\bm{r},t)&=
        \Delta_{\ell,v}(\bm{r})-\frac{\hbar^2}{2m^*}\left(\delta\bm{k}_\xi^\ell+\frac{e\bm{A}}{\hbar}\right)^2\\
        &=\Delta_{\ell,v}(\bm{r})-\frac{\hbar^2}{2m^*}\left[(\delta\bm{k}_\xi^\ell)^2+\frac{e^2 A_0^2}{\hbar^2}+2\delta\bm{k}_\xi^\ell\cdot\frac{e\bm{A}}{\hbar}\right]\\
        &=\Delta_{\ell,v}(\bm{r})-\frac{\hbar^2}{2m^*}\Big[(\delta\bm{k}_\xi^\ell)^2+\frac{e^2 A_0^2}{\hbar^2}\\
        &+\delta{k}_{\xi,x}^\ell \frac{e A_{0}}{\hbar}(e^{i\Omega t}+e^{-i\Omega t})+i\delta{k}_{\xi,y}^\ell \frac{e A_{0}}{\hbar}(e^{i\Omega t}-e^{-i\Omega t})\Big],
    \end{aligned}
\end{equation}
with Fourier components 
\begin{subequations}
    \begin{align}
        H_{0}&=\mathcal{H}_{\text{kin}}-\frac{e^2 A_0^2}{2m^*}
        \mathds{1}_{2}^{\ell}\otimes\mathds{1}_{2}^{\xi},\\
        H_{1}&=-\frac{e\hbar A_0}{2m^*}(\delta{k}_{\xi,x}^\ell +i\delta{k}_{\xi,y}^\ell)
        \mathds{1}_{2}^{\ell}\otimes\mathds{1}_{2}^{\xi},\\
        H_{-1}&=H_{1}^{\dagger}.
    \end{align}
\end{subequations}
Using the Magnus expansion, we obtain 
\begin{subequations}
    \begin{align}
        H_{\text{eff}}^{(0)}&=H_{0}=\mathcal{H}_{\text{kin}}-\frac{e^2 A_0^2}{2m^*}
        \mathds{1}_{2}^{\ell}\otimes\mathds{1}_{2}^{\xi},\\
        H_{\text{eff}}^{(1)}&=\frac{1}{\hbar\Omega}[H_{1},H_{-1}]=0,\\
        H_{\text{eff}}^{(2)}&=\frac{1}{2(\hbar\Omega)^2}[H_{1},[H_{0},H_{-1}]]+h.c.\notag\\
        &=-\left[\frac{e A_0}{2m^*\Omega}(\bm{\kappa}_{t,\xi}-\bm{\kappa}_{b,\xi})\right]^2
        \left\{T_{\xi}(\bm{r})\left(\frac{\gamma_x+i\gamma_y}{2}\right)+h.c.\right\}\otimes\mathds{1}_2^{\xi},
    \end{align}
\end{subequations}
where $\mathds{1}_2^\ell$ and $\mathds{1}_2^\xi$ are $2\times 2$ identity matrices in the layer and valley spaces, respectively, and $\gamma_i$ ($i=x,y,z$) are Pauli matrices in the layer space. Combining these terms, we obtain 
\begin{equation}
    H_{\text{eff}}=h_{\ell,\xi}^{'}(\bm{k},\bm{r})\otimes\mathds{1}_2^{\ell}\otimes\mathds{1}_2^{\xi}
    +\left\{\Delta_{T}^{\xi}(\bm{r})\left(\frac{\gamma_x+i\gamma_y}{2}\right)
    +h.c.\right\}\otimes\mathds{1}_2^{\xi},
\end{equation}
with 
\begin{subequations}
    \begin{align}
        h_{\ell,\xi}^{'}(\bm{k},\bm{r})&=
        -\frac{\hbar^2}{2m^*}\{(\delta\bm{k}_\xi^\ell)^2+\frac{e ^2 A_0^2}{\hbar^2}\}+\Delta_{\ell,v}(\bm{r}),\\
        \Delta_{T}^{\xi}(\bm{r})&=T_\xi (\bm{r}) \left\{ 1-\left[\frac{e A_0}{2m^*\Omega}(\bm{\kappa}_{t,\xi}-\bm{\kappa}_{b,\xi})\right]^2\right\}.
    \end{align}
\end{subequations}
Therefore, the leading effect of high-frequency CFL driving within the free-electron framework is a constant quasienergyshift $-\frac{e^2 A_0^2}{2m^*}$ which is independent of the driving frequency and the valley. 
There is no time-reversal symmetry breaking feature in this shift. As the $1/\Omega$ correction vanishes, the time-reversal symmetry breaking effect is also absent at this order. This is in striking contrast to the result that is obtained by starting from the Dirac model and only applying the free-electron approximation at the end.

\end{document}